%% file: aaskaii_template.tex
\documentclass[a4paper,11pt]{article}
\usepackage{aaskaiid}
\usepackage{ragged2e}
\usepackage{multirow}
\usepackage{orcidlink}

\newcommand\arcdeg{\mbox{$^\circ$}}%
\newcommand\arcmin{\mbox{$^\prime$}}%
\newcommand\arcsec{\mbox{$^{\prime\prime}$}}%

\title{The Physics of the Cold and Warm Interstellar Medium Through Radio Recombination Lines}
\ShortTitle{RRLs and the Diffuse ISM}

\author[1]{Pedro Salas\orcidlink{0000-0001-8271-0572}}
\ShortName{Salas et al.} 
\author[1]{Kimberly L. Emig\orcidlink{0000-0001-6527-6954}}
\author[2]{Matteo Luisi\orcidlink{0000-0001-8061-216X}}
\author[3,4]{Antoine Gusdorf\orcidlink{0000-0002-0354-1684}}
\author[3,4]{Lucie Cros\orcidlink{0000-0003-4422-8720}}
\author[5]{Loren D. Anderson\orcidlink{0000-0001-8800-1793}}
\author[6]{Alexander Tielens\orcidlink{0000-0003-0306-0028}}

\affiliation[1]{National Radio Astronomy Observatory, USA}
\affiliation[2]{Westminster College, USA}
\affiliation[3]{Laboratoire de Physique de l’École Normale Supérieure, ENS, Université PSL, CNRS, Sorbonne Université, Université Paris Cité, F-75005, Paris, France}
\affiliation[4]{LUX, Observatoire de Paris, Université PSL, Sorbonne Université, 75014 Paris, France}
\affiliation[5]{West Virginia University, USA}
\affiliation[6]{Leiden Observatory, Leiden University, Leiden, The Netherlands}

\abstract{
The interplay between star formation and the interstellar medium (ISM) gives rise to a feedback cycle that is one of the key drivers of galaxy evolution. The coldest gas in the ISM provides the fuel for star formation, and, in turn, the ISM is affected by feedback in the form of photons, shocks, and cosmic rays. Radio recombination lines (RRLs) offer a unique way of characterizing the gas in the ISM as extinction free probes of the gas kinematics, density, temperature, and radiation field. Hydrogen RRLs (HRRLs) probe gas where hydrogen is ionized, while carbon RRLs (CRRLs) probe cold gas in the HI-to-H$_{2}$ transition layers. The superb sensitivity, spatial resolution, large field of view, and frequency resolution and coverage of the Square Kilometre Array (SKA) will allow for efficient observations of RRLs, that will make significant impacts to the study of the diffuse ISM. By observing RRLs with the SKA we will be able to determine the thermal and turbulent balance, chemical enrichment, sizes, and ionization rate of the cold and ionized gases from degree-scales down to arcsecond scales in our Galaxy.
}

\begin{document}
\maketitle

\section{Introduction}

The interstellar medium (ISM) is a key component of galaxy evolution, as the fuel for star formation and repository of stellar ejecta. Feedback in the form of shocks, photons and cosmic rays, enriches, heats, ionizes, dissociates, stirs, and carves the ISM. In turn, the ISM cools, flows, transforms, expands and contracts, leading to the formation of new stars and/or the quenching of star formation. Stars are formed in the densest parts of the ISM. Hot massive stars create pockets of ionized gas in HII regions. Some of the Lyman continuum photons are able to escape the boundaries of HII regions and ionize the surrounding gas, while less energetic photons couple to the gas through photoelectrons in photodissociation regions (PDRs). Supernovae explosions carve out bubbles that can extend several parsecs, thus providing pathways for Lyman continuum photons to escape the disk of star forming galaxies and reach their halos. Supernovae explosions, stellar winds, jets and outflows are sources of turbulence, which provides the ISM with support against gravity, and they also shock the gas, modifying its ionization, temperature, density and dynamics. Turbulence also gives the ISM an inhomogeneous structure, filamentary and fractal, thus further providing transparent pathways for ionizing photons to travel large distances. Stellar ejecta enrich the ISM with the products of nucleosynthesis. Throughout the cycle from gas to stars, the gas in the ISM goes through several states; hot and ionized, warm and ionized, warm and neutral, and cold and neutral. Characterizing the physical properties of interstellar gas, with spatially and spectrally resolved observations, will help us understand the role of the ISM in the evolution of galaxies and the details about star formation.

In this chapter we describe how the study of the diffuse ISM, cold and neutral, and warm and ionized, in our Galaxy will be enhanced by observations of radio recombination lines (RRLs) with the Square Kilometre Array (SKA).

\subsection{Radio Recombination Lines}

RRLs are the spectral lines produced by the cascade of an electron from energy levels with principal quantum numbers $n\gtrsim30$ after it is captured by an ion. Given their long wavelengths they are not attenuated by dust, making them extinction free probes of the ISM. Moreover, since the rest frequencies of the transitions are well known, from the Rydberg formula \citep{Rydberg1890}, RRLs also probe the kinematics of interstellar gas. Their brightness can be a sensitive probe of the gas physical conditions (e.g., emission measure, density and temperature) \citep[e.g.,][]{Oonk2017,Emig2020,Roshi2025,Cros2025}. They have a Zeeman splitting of $2.8$~Hz~$\mu$G$^{-1}$ \citep{Greve1980}, making them potential magnetometers. The emission from RRLs has been described by \citet{Shaver1975}, \citet{Salgado2017b} and \citet{Prozesky2018}.

RRLs from hydrogen (HRRLs) and helium (HeRRLs), mainly probe the warm ionized gas in the ISM. Under typical conditions their line widths are ${\approx}25$~km~s$^{-1}$ \citep[e.g.,][]{Anderson2021}. RRLs from carbon (CRRLs) and sulfur (SRRLs) probe cold gas in the HI-to-H$_{2}$ transition layers of the ISM \citep[e.g.,][]{Salas2018,Cuadrado2019,Pabst2024}.

A summary of recent and on-going surveys of the Galactic ISM using RRLs is presented in Table~\ref{tab:rrl_surveys}. Of these, only two have sub-arcminute resolution (THOR and GLOSTAR, the only interferometric surveys in Table~\ref{tab:rrl_surveys}), and only one is probing RRLs below $1$~GHz (GDIGS-Low). As we will see, the SKA has the potential to transform the study of the diffuse ISM by enabling observations at sub-arcminute resolution with a sensitivity comparable to that achieved by single dish surveys.

To determine the sensitivity of a survey in terms of emission measure we use Eq.~(A6) of \citet{Anderson2021}, and the signal-to-noise prescription of \citet{Lenz1992} assuming a full width at half maximum of $25$~km~s$^{-1}$. The EM values quoted in Table~\ref{tab:rrl_surveys} are for a $3\sigma$ sensitivity. We also note that at frequencies ${\lesssim}400$~MHz the line brightness becomes a stronger function of the electron density and radiation field, so the EM sensitivity can change depending on these values, being more sensitive to lower density gas.

\begin{table}[h]
    \centering
    \caption{Recent and on-going RRL surveys}
    \label{tab:rrl_surveys}
    \begin{tabular}{lccccccccc}
    \hline
        Survey & rms & rms & $\Delta\varv$ & $\theta$ & $\nu$ & area & EM & Ref \\
         & (mK) & (mJy beam$^{-1}$) & (km s$^{-1}$) & ($\arcmin$) & (GHz) & (sq. deg.) & (pc cm$^{-6}$) & \\
        \hline
         THOR & 814 & 2.3 & 10 & 0.7 & 1.4 & 104 & $9.9\times10^{4}$ & (5),(6)\\
         GLOSTAR & 206 & 3 & 5 & 0.42 & 5.3 & 145 & $6.6\times10^{4}$ & (7) \\
         GDIGS & 10 & 5 & 0.5 & 2.65 & 5.76 & 35 & 1100 & (1) \\
         HIPASS-RRL & 4.5 & 6.4 & 20 & 14.4 & 1.4 & 2160 & 760 & (2) \\
         SIGGMA & 6.5 & 0.65 & 5.1 & 6 & 1.35 & 114$^{\ddagger}$ & 540 & (3) \\
         GDIGS-Low 340 & 60$^{\dagger}$ & 30$^{\dagger}$ & 0.5 & 48 & 0.34 & 70 & 400$^{\dagger}$ & (8) \\
         FAST-GPPS & 6.3 & 0.25 & 2.2 & 3 & 1.24 & 2720 & 310 & (4) \\
         GDIGS-Low 800 & 18$^{\dagger}$ & 9$^{\dagger}$ & 0.5 & 18 & 0.79 & 70 & 280$^{\dagger}$ & (8) \\
    \hline
    \end{tabular}
    
    \bigskip
    \RaggedRight
    Notes
    (1) \citet{Anderson2021}
    (2) \citet{Alves2015}
    (3) \citet{Liu2019}
    (4) \citet{Hou2022}
    (5) \citet{Beuther2016}
    (6) \citet{Wang2020}
    (7) \citet{Brunthaler2021}
    (8) Salas et al. in prep.
    $^{\ddagger}$ Inner Galaxy data release.
    $^{\dagger}$ Expected sensitivity.
\end{table}

\subsection{The Warm Ionized Medium}

The warm ionized medium (WIM) is a major component of the Milky Way. It has a volume filling factor of ${\approx}20\%$ and accounts for $90\%$ of all the ionized hydrogen in the Galaxy. It is characterized by an ionization parameter $10^{-4}$ to $10^{-3}$; about 1 to 2 orders of magnitude smaller than for classical HII regions \citep{Haffner2009}. It has a scale height of $1$ to $1.9$~kpc, and it is mainly photoionized. In terms of its physical properties, it spans densities from ${\sim}0.01$~cm$^{-3}$ at high Galactic latitudes, to ${\sim}100$~cm$^{-3}$ in the vicinity of HII regions, the bulk of which are in the Galactic disk. Its temperature has been estimated to be between $6000$ and $10000$~K \citep{Madsen2006}. In other galaxies the WIM is referred to as diffuse ionized gas (DIG). For a review of the WIM see \citet{Haffner2009}. Sudies of HII regions with the SKA are presented in a separate chapter.

The Galactic halo is ionized by the photons produced in the Galactic disk, however, the details of how these photons travel through the ISM to reach the halo are still not well understood. To escape the Galactic disk these photons must travel through transparent pathways, chimneys or ``worms''. Turbulence \citep[e.g.,][]{Wood2005} and large bubbles, created by clustered supernovae explosions, could produce these pathways. For these mechanisms to explain the prevalence of the Galactic halo, the number of such transparent pathways must be large enough that enough photons can escape to the halo. Large scale mapping of the WIM, and the ISM in general, could be used to count the number of such structures, and their morphology and kinematics to determine their origin. Moreover, if we are also able to determine the ionization state of the ISM, we could quantify how many photons are able to escape through the identified pathways.


The WIM has been studied through a variety of probes. The main one being through the H$\alpha$ recombination line at optical wavelengths, and including pulsar dispersion measures, thermal continuum, radio recombination lines, far-infrared (FIR) fine structure lines, and mid-infrared (MIR) lines. All of these probe different regimes in terms of their emission measure (EM) sensitivity (or dispersion measure in the case of pulsars). For example, H$\alpha$ observations probe down to EM${\sim}1$~pc~cm$^{-6}$ on degree scales \citep{Haffner2010}, MIR lines ${\sim}10$~pc~cm$^{-6}$ on arcsecond scales \citep{Kulkarni2024}, and HRRLs ${\approx}300$~pc~cm$^{-6}$ \citep{Hou2022} on $3\arcmin$ scales. This has left gaps in our ability to connect the density components and physical scales probed by different tracers. Being able to link some of these components would help a better characterization of the morphology of the WIM. An order of magnitude leap in sensitivity and two orders of magnitude in angular resolution, with respect to current RRL surveys (Table~\ref{tab:rrl_surveys}), are required to connect observations of HRRLs to those in the mid-IR with JWST. Carrying out observations with these improvements would increase the coverage of emission measures and angular scales probed, resulting in a more comprehensive census of the gas conditions present in the Galactic WIM.

\subsection{Cold Neutral Gas}

Cold gas is the fuel for star formation. Stars are formed in dense pockets of cold gas, embedded in molecular clouds. This dense cold gas plays a central role in the baryonic cycle, and understanding how molecular clouds form and evolve is a subject of active study \citep[e.g.,][]{Dobbs2014,Chevance2023}. The formation of molecular clouds requires high enough densities that the rate of formation of molecular hydrogen, the most abundant molecule in the Universe, can outpace its destruction. Direct observation of H$_{2}$ is possible through observations in the ultraviolet of its electronic transitions in absorption studies, or in conditions where the gas is sufficiently warm ($T\gtrsim500$~K) that its rotational transitions are excited. This makes H$_{2}$ invisible under most of the conditions critical for star formation in our Galaxy. So, to study the formation and evolution of molecular clouds requires alternative probes.

At radio wavelengths the rotational transitions of CO have been the main tool used for the study of molecular clouds, while the hyperfine transition from atomic hydrogen has been used to study cold atomic gas. However, a significant gap is left by the use of these tracers, as they are not good probes of the regions where H$_{2}$ is being formed. The formation of CO requires the presence of H$_{2}$, thus it probes regions where hydrogen is already predominantly molecular. On the other end, HI probes a variety of gas conditions, from diffuse warm gas to the cold hydrogen forming molecular clouds. This complicates the interpretation of 21~cm-HI observations as probes of the HI-to-H$_{2}$ transition layer. Other, fainter, tracers include 18~cm-OH \citep[e.g.,][]{Busch2021}, CH \citep[e.g.,][]{Jacob2021} and carbon RRLs \citep[CRRLs, e.g.,][]{Emig2025}.

The HI-to-H$_{2}$ transition takes place in regions where carbon is ionized, as carbon has a lower ionization potential ($11.3$~eV) than that of hydrogen ($13.6$~eV). Moreover, the brightness of CRRLs is inversely proportional to the gas temperature, so they are brighter towards regions at colder temperatures (${\lesssim}1000$~K). This makes CRRLs a direct probe of the HI-to-H$_{2}$ transition. Observations of CRRLs with the SKA will provide key insights into the lifecycle of molecular clouds, the role of turbulence in the HI-to-H$_{2}$ transition layers, and, in combination with observations of other tracers, into the physical and chemical structures of diffuse PDRs.

\section{SKA Observations of the Cold and Warm ISM through RRLs}

The SKA will provide unique capabilities to carry out RRL observations at sub-arcminute resolution, spanning a wide frequency range, and with unprecedented sensitivity. In this section we explore what we can expect from Galactic RRL observations with the SKA array assembly * (AA*) and AA4. We consider the following cases, (i) observations with high surface-brightness sensitivity, at the expense of angular resolution, (ii) observations at higher angular resolution, at the expense of surface-brightness sensitivity, and (iii) high spectral resolution observations to target narrow lines.

We present common parameters for the SKA-Mid and SKA-Low in Table~\ref{tab:ska_pars}. To evaluate the field of view (FoV) we use $\theta=1.2\lambda/D$, with $\lambda$ the observing wavelength and $D$ the antenna/station diameter. We evaluate the FoV at the highest frequency in a band, and adopt the largest antenna/station size (15 m for SKA-Mid and 40 m for SKA-Low). For the angular and spectral resolution, we consider the values attained at the lowest frequency end of a band, since in post processing we will want to stack RRLs, thus we will be limited to the properties of the lowest frequency transition in one band.

\begin{table}[h]
	\centering
	\caption{SKA common parameters}
	\label{tab:ska_pars}
	\begin{tabular}{lcccccccc}
		\hline
		Band & SA & $\nu_{\mathrm{c}}$ & $\Delta\nu$ & FoV & $\Delta\nu$ & $\Delta\varv$ & \# $\alpha$ lines & \# $\alpha$ lines zoom \\
                 &    & (GHz) & (MHz) & ($\arcmin$) & (kHz) & (km s$^{-1}$) & \\
		\hline
        \multicolumn{9}{c}{SKA-Low}\\
        \hline
        & & 0.2 & 300 & 56 & 5.43 & 4.6--32.4 & 243 & 27 \\
        \hline
        \multicolumn{9}{c}{SKA-Mid}\\
        \hline
		1 & 15-m & 0.7 & 700 & 78 & 13.4 & 3.8--11.5 & 82 & 10 \\
		2 & & 1.31 & 810 & 46.8 & 13.4 & 2.3--4.2 & 36 & 9 \\
		5a & 15-m & 6.55 & 3900 & 9.7 & 53.8 & 1.9--3.5 & 21 & 4 \\
		\hline
	\end{tabular}
    
    \bigskip
    \RaggedRight
    SKA-Mid band, subarray, central frequency, instantaneous bandwidth, field of view at the highest frequency, channel width across full bandwidth, velocity resolution, number of $\alpha$ RRLs in the band (for H, C, He or S), number of $\alpha$ RRLs (for H, C, He or S) using a zoom mode with $\Delta v{\approx}1$~km~s$^{-1}$ around $\nu_{\mathrm{c}}$.
    For SKA-Mid it will be possible to use up to $4$ zoom windows.
    For SKA-Low there will be up to $64$ zoom windows.
\end{table}

We note that at the low end of SKA-Mid band 1, and below ${\approx}163$~MHz for the SKA-Low, the spectral resolution is coarser than $10$~km~s$^{-1}$, which is close to the $27.4$~km~s$^{-1}$ velocity separation between the RRLs from carbon and helium (HeRRLs). At these frequencies, observations of CRRLs and/or HeRRLs would require the use of a zoom mode, given the current baseline design.

To evaluate the contribution from the Galactic continuum to the system temperature we use the built in model of the SKA sensitivity calculator. To estimate the sensitivity of the SKA for Galactic RRL observations we employ this model of the Galactic emission and adopt the coordinates of the Galactic center as representative ($(\alpha,\delta)_{\rm{J2000}}=(17^{\rm{h}}45^{\rm{m}}37.\!\!^{\rm s}2159, -28\arcdeg56\arcmin10.\!\!^{\prime\prime}184)$). For the SKA-Mid results we adopt an elevation of $45\arcdeg$, and for the SKA-Low a minimum elevation of $20\arcdeg$.

For bands 1 and 5a we present calculations using only the 15-m antennas, as they are equipped to cover the full bandwidth of each band, thus maximizing the number of RRLs that can be observed simultaneously. We use Briggs weighting with a robust factor of 1, and for the SKA-Mid we include a taper to increase the surface-brightness sensitivity to extended emission. We use the continuum calculator, and we present the weighted spectral sensitivity (rms columns in Tables~\ref{tab:aa*_ska_rrl_obs} and \ref{tab:aa4_ska_rrl_obs}), ignoring the contribution from confusion to the noise budget.

In Table~\ref{tab:ska_pars} we have that the number of $\alpha$ lines is $\geq21$, however some of these lines will be lost due to radio frequency interference (RFI), poor data quality, and/or the instrumental response (e.g., spectral window edges). Given the uncertainty in the actual number of lines available for stacking, we present our calculations per line, and assuming that $70\%$ of the lines will be available for stacking (rms$_{0.7}$ columns in Tables~\ref{tab:aa*_ska_rrl_obs} and \ref{tab:aa4_ska_rrl_obs}). We assume that the noise will be reduced by the square root of the number of lines.  

\subsection{AA$^{*}$}

In Table~\ref{tab:aa*_ska_rrl_obs} we present the expected sensitivity of the SKA AA$^{*}$ for RRL observations. 

\begin{table}[h]
	\centering
	\caption{RRL Observations with the SKA AA$^{*}$}
	\label{tab:aa*_ska_rrl_obs}
	\begin{tabular}{lcccccccc}
		\hline
		Band & SA & taper & $\theta$ & $\Delta t$ & rms & rms & rms$_{0.7}$ & EM \\
        & & ($\arcsec$) & ($\arcmin$) & (hours) &(mJy beam$^{-1}$) & (mK) & (mK) & (pc cm$^{-6}$)\\
	\hline
        \hline
        \multicolumn{9}{c}{High Brightness Sensitivity}\\
        \hline
        \hline
        \multicolumn{9}{c}{SKA-Low}\\
        \hline
        ---  & inner 350 m & --- & 7.2 & 8 & 3.8 & 635 & 49 & 1285 \\
        \hline
        \multicolumn{9}{c}{SKA-Mid}\\
        \hline
		1  & 15-m & 2048 & 23  & 8 & 60  & 76 & 10 & 541 \\
		2  &      & 1058 & 12  & 8 & 16  & 83 & 17 & 1141\\
		5a & 15-m & 220  & 2.5 & 8 & 4.8 & 12 & 3.2 & 990\\
	\hline
        \hline
        \multicolumn{9}{c}{High Angular Resolution}\\
        \hline
        \hline
        \multicolumn{9}{c}{SKA-Low}\\
        \hline
        ---  & inner 3 km & --- & 3.75 & 8 & 2.3 & 1390 & 107 & 2908 \\
        \hline
        \multicolumn{9}{c}{SKA-Mid}\\
        \hline
        1  & 15-m & 128 & 2.5  & 8 & 6.45 & 815 & 108 & 5825 \\
        2  &      & 66  & 1.34 & 8 & 1.7  & 213 & 43 & 2928 \\
        5a & 15-m & 14  & 0.27 & 8 & 0.52 & 65  & 17 & 5365 \\
        \hline
	\end{tabular}
    
    \bigskip
    \RaggedRight
    SKA-Mid band, subarray, taper, synthesized beam size at the lowest frequency, integration time, sensitivity, surface brightness sensitivity, surface brightness sensitivity after stacking $70\%$ of the $\alpha$ RRLs, $3\sigma$ emission measure sensitivity.
\end{table}

As an example, we estimate how long it would take to reproduce a survey like the Green Bank Telescope (GBT) Diffuse Ionized Gas Survey GDIGS \citep{Anderson2021} using band 5a of the AA$^{*}$ for the high-brightness sensitivity case. We use GDIGS for comparison as it represents the state of the art in the field, and it was a dedicated RRL survey. GDIGS fully sampled the mapped area of ${\approx}37$ sq. deg. It would require roughly $451$ pointings to cover the same area with 15-m antennas. To match the surface-brightness sensitivity of GDIGS ($10$~mK), it takes ${\approx}0.82$~hours of AA$^{*}$ time, so it would take $370$ hours to cover the same area to a similar depth, ignoring overheads. GDIGS required $380$~hours of time to be completed (including overheads and calibration). Thus, with the AA$^{*}$ it would be possible to extend a survey like GDIGS towards lower declinations, while simultaneously providing a slight increase in angular resolution.


The high angular resolution observations would provide more than an order of magnitude improvement in sensitivity with respect to existing interferometric RRL surveys. This will be transformative, as it will enable mapping the morphology and kinematics of the ISM gases probed by RRLs. For example, the warm ionized gas filaments studied by \citet{Emig2022} had a median EM of $5200$~pc~cm$^{-6}$ at an angular resolution of $2\arcmin$.

\subsection{AA4}

\begin{table}[th]
	\centering
	\caption{RRL Observations with the SKA AA4}
	\label{tab:aa4_ska_rrl_obs}
	\begin{tabular}{lcccccccc}
		\hline
		Band & SA & taper & $\theta$ & $\Delta t$ & rms & rms & rms$_{0.7}$ & EM \\
        & & ($\arcsec$) & ($\arcmin$) & (hours) &(mJy beam$^{-1}$) & (mK) & (mK) & (pc cm$^{-6}$)\\
	\hline
        \hline
        \multicolumn{9}{c}{High Brightness Sensitivity}\\
        \hline
        \hline
        \multicolumn{9}{c}{SKA-Low}\\
        \hline
          & inner 350 m & & 7.2 & 8 & 3.8 & 635 & 49 & 1285 \\
        \hline
        \multicolumn{9}{c}{SKA-Mid}\\
        \hline
		1  & 15-m & 2048 & 16  & 8 & 22  & 60 & 8 & 427 \\
		2  &      & 1058 & 8.3 & 8 & 5.9 & 17 & 3.4 & 233 \\
		5a & 15-m & 220  & 1.7 & 8 & 1.8 & 5  & 1.3 & 413 \\
	\hline
        \hline
        \multicolumn{9}{c}{High Angular Resolution}\\
        \hline
        \hline
        \multicolumn{9}{c}{SKA-Low}\\
        \hline
          & inner 3 km & --- & 2 & 8 & 1.6 & 3400 & 260 & 7859 \\
        \hline
        \multicolumn{9}{c}{SKA-Mid}\\
        \hline
        1  & 15-m & 128 & 3.2  & 8 & 3.4  & 235 & 31 & 1675 \\
        2  &      & 66  & 1.7  & 8 & 0.9  & 63  & 13 & 866 \\
        5a & 15-m & 14  & 0.34 & 8 & 0.27 & 20  & 5.3 & 1650 \\
        \hline
	\end{tabular}
    
    \bigskip
    \RaggedRight
    SKA-Mid band, subarray, taper, synthesized beam size at the lowest frequency, integration time, sensitivity, surface brightness sensitivity, surface brightness sensitivity after stacking $70\%$ of the $\alpha$ RRLs, $3\sigma$ emission measure sensitivity.
\end{table}

The array assembly 4 (AA4) baseline design for the SKA is described in \citet{Braun2019}. In Table~\ref{tab:aa4_ska_rrl_obs} we present the expected sensitivity of the SKA AA4 for RRL observations. In general, the SKA-Mid AA4 would provide a factor of ${\approx}2$ improvement in sensitivity and ${\approx}1.5$ in angular resolution with respect to the AA* for observations at high brightness sensitivity. At high angular resolution, the SKA-Mid AA4 would deliver a factor of $3$ improved sensitivity than AA$^{*}$, at a slightly coarser angular resolution.

\section{Beyond the Baseline}

With the current baseline design for the SKA, the spectral resolution of the continuum mode is not high enough to spectrally resolve narrow RRLs (widths ${\approx}1$~km~s$^{-1}$), such as those probing cold gas from either carbon or hydrogen \citep[][]{Emig2020}. Thus, their observation requires the use of zoom windows over a smaller bandwidth. These observations would benefit from higher spectral resolution without having to restrict the bandwidth. A high spectral resolution mode ($\Delta v\approx1$~km~s$^{-1}$) over the full bandwidth would improve the sensitivity for narrow line observations by a factor of 2 to 3 (at a $1$~km~s$^{-1}$ spectral resolution). Alternatively, more zoom windows could be offered, so that more RRLs can be observed simultaneously at high spectral resolution.

\section{Summary}

In summary, the SKA will provide and unprecedented view of the diffuse ISM in our Galaxy through RRL observations thanks to its sensitivity, spatial resolution and frequency coverage. This will enable mapping the kinematics and morphology of the diffuse ionized and cold gases, and also the determination of their physical properties (e.g., density for ionized and cold gases, and temperature for cold gases). These capabilities will have a long lasting impact in studies of the cold and ionized components of the diffuse ISM. For the WIM, the SKA will enable building a model of the distribution, kinematics and density of the WIM, which can be interpreted in its relation to star formation and the feedback processes that influence the ISM. For cold neutral gas the SKA will provide key information about how molecular clouds form, what are the roles of turbulence versus large scale dynamics in the lifecycle of molecular clouds, and what are the physical and chemical structures of cold neutral clouds.

\include{journal-names}
\bibliographystyle{abbrvnat-maxbibnames4}
\bibliography{chapter}

\end{document}

%% file: journal-names.tex
\newcommand{\actaa}{Acta Astron.} 
\newcommand{\araa}{ARA\&A} 
\newcommand{\aar}{A\&ARv} 
\newcommand{\aapr}{A\&ARv} 
\newcommand{\ab}{Astrobiol.} 
\newcommand{\aj}{AJ} 
\newcommand{\apj}{ApJ} 
\newcommand{\apjl}{ApJL} 
\newcommand{\apjs}{ApJSS} 
\newcommand{\ao}{Appl. Opt.} 
\newcommand{\apss}{Astro. \& Space Sci.} 
\newcommand{\aap}{A\&A} 
\newcommand{\aaps}{A\&AS.} 
\newcommand{\baas}{Bull. Am. Astron. Soc.} 
\newcommand{\caa}{Chinese A\&A} 
\newcommand{\cjaa}{Chinese J. A\&A} 
\newcommand{\cqg}{Class. Quantum Gravity} 
\newcommand{\gal}{Galaxies} 
\newcommand{\gca}{Geo. Cosmo. Acta} 
\newcommand{\icarus}{Icarus} 
\newcommand{\jcap}{JCAP} 
\newcommand{\jgr}{J. Geophys. Res.} 
\newcommand{\jgrp}{J. Geophys. Res. Planets} 
\newcommand{\jqsrt}{J. Quant. Spectrosc. Radiat. Transf.} 
\newcommand{\memsai}{Mem. SAIt} 
\newcommand{\mnras}{MNRAS} 
\newcommand{\nat}{Nature} 
\newcommand{\nastro}{Nat. Astron.} 
\newcommand{\ncomms}{Nat. Commun.} 
\newcommand{\nphys}{Nat. Phys.} 
\newcommand{\na}{New Astron.} 
\newcommand{\nar}{New Astron. Rev.} 
\newcommand{\physrep}{Phys. Rep.} 
\newcommand{\pra}{Phys. Rev. A} 
\newcommand{\prb}{Phys. Rev. B} 
\newcommand{\prc}{Phys. Rev. C} 
\newcommand{\prd}{Phys. Rev. D} 
\newcommand{\pre}{Phys. Rev. E} 
\newcommand{\prx}{Phys. Rev. X} 
\newcommand{\prl}{Phys. Rev. Let.} 
\newcommand{\psj}{Planet. Sci. J.} 
\newcommand{\planss}{Planet. Space Sci.} 
\newcommand{\pnas}{Proc. Natl Acad. Sci. USA} 
\newcommand{\procspie}{Proc. SPIE} 
\newcommand{\pasa}{PASA} 
\newcommand{\pasj}{PASJ} 
\newcommand{\pasp}{PASP} 
\newcommand{\rmxaa}{RMXAA} 
\newcommand{\sci}{Science} 
\newcommand{\sciadv}{Sci. Adv.} 
\newcommand{\solphys}{Sol. Phys.} 
\newcommand{\sovast}{Soviet Ast.} 
\newcommand{\ssr}{Space Sci. Rev.} 
\newcommand{\uni}{Universe} 